# Oscillating dipole model for the X-ray standing wave enhanced fluorescence in periodic multilayers

Jean-Michel André[*], Karine Le Guen, Philippe Jonnard

*Laboratoire de Chimie Physique – Matière et Rayonnement, UPMC Univ Paris 06, CNRS UMR 7614, 11 rue Pierre et Marie Curie, F-75231 Paris cedex 05, France*

*Abstract*

Periodic multilayers give rise to enhanced X-ray fluorescence when a regime of standing waves occurs within the structure. This regime may concern the primary radiation used to induce the fluorescence, the secondary radiation of fluorescence or both of them. Until now, existing models only dealt with standing wave regime of primary radiation. We present a theoretical approach based on the oscillating dipole model and the coupled-wave theory that can treat efficiently any standing wave regime. We compare our simulations to experimental data available in the literature.

*(\*) Corresponding author*

Dr. Jean-Michel André

jean-michel.andre1@upmc.fr

Laboratoire de Chimie Physique – Matière et Rayonnement

11 rue Pierre et Marie Curie, F-75231 Paris cedex 05, France

Tel/Fax : (33) 1 44 27 66 14 / (33) 1 44 27 62 26

# 1. Introduction

X-ray Bragg diffraction by a periodic multilayer structure gives rise to a system of X-ray standing waves (XSW) that can be profitably used to determine the microstructural properties of the stack [1–6]; by adjusting the grazing angle $\theta_0$ in the Bragg domain, it is possible to localize the peaks of electric field intensity within the structure in the regions of interest for the analysis. The primary incident field can generate photoelectrons, fluorescence emission and be elastically (Rayleigh) or inelastically (Raman-Compton) scattered. The production of secondary X-rays makes it possible to probe the structure by different ways: as mentioned in reference [5], XSW enhanced fluorescence is rather interesting to analyse high-Z layers since photoelectric cross sections scales as $Z^4$ while the elastic or inelastic scattering profiles are more sensitive to the cross sections of low Z materials. Modelling these emissions generally consists in:

- calculating at a given depth $z$ the intensity $I_{exc}(z, \theta_0, E_0)$ of the local exciting electric field resulting from the interferences between the incident and reflected waves,
- then considering an exponentially attenuation of the secondary emitted X-rays (Beer-Lambert law)
- and finally performing an integration along the depth of the structure; the result is weighted by the cross section $\tau(E_0)$ of the phenomenon which strongly depends on the photon energy.

This can be summarized by the formula:

$$I(\theta, E) = \tau(E_0) \int_0^L I_{exc}(z, \theta_0, E_0) \exp\left[-\mu(E) \frac{z}{\sin(\theta)}\right] dz$$

(1)

In Eq. (1), $E_0$ is the energy of the primary photon, $\theta_0$ the grazing angle, $E$ is the energy of the secondary photon, $\theta$ the take-off angle, µ the linear absorption coefficient for the secondary radiation and $L$ the thickness of the multilayer structure. $I_{exc}(z, \theta_0, E_0)$ is computed according to standard techniques but the generalized recursive Parratt method [7,8] is often implemented. This model can be refined to take into account some particular effects such as the inhomogeneity of the wave in absorbing media, secondary fluorescence, roughness [1,4]. This kind of theoretical approach seems to be valid as long as the secondary emitted wave does not encounter Bragg diffraction.

When the secondary emitted X-rays are Bragg diffracted, this approach is no longer relevant; such an experiment in fluorescence mode has been carried with the Fe Kα fluorescence line emitted from Fe/C multilayers excited by the Cu Kα line [2,3]. This configuration is similar to the one reported in experiments by Kossel *et al.* [9] then by Jonnard *et al.* with multilayers excited by an electron beam [10,11]. Kossel-like experiments can be interpreted by means of Lorentz reciprocity theorem [2,3] as initially proposed by Laue [12].

We propose here a more direct approach that can be applied for fluorescence, elastic and inelastic scattering, both for primary and secondary radiations undergoing Bragg diffraction. The idea is:

- to consider the sources of secondary radiation as oscillating dipoles radiating at the frequency of the secondary radiation, excited by the local electric field resulting from the primary radiation,
- then to calculate the propagation of the total field satisfying a second-order differential equation with the appropriate boundary conditions and finally to determine the intensity in far-field of the secondary radiation.

The total field is the sum of the homogeneous field satisfying the propagation equation without second member plus a source field, which is a particular solution of the propagation equation with a second source term given by the current density induced by the dipole. The propagation problem can be solved by means of the dyadic Green function formalism [13,14], but since this approach requires a rather high level in mathematics, we prefer using a more direct manner of solving the differential inhomogeneous equation implementing partial Fourier transform.

Section 2 is devoted to the theoretical development: Section 2.1 treats the radiation of an oscillating dipole from a periodic multilayer structure; Section 2.2 presents a calculation of the local exciting electric field resulting from the primary radiation by the coupled-wave theory; Section 2.3 deals with the calculation of the induced dipole and its distribution within the stack; Section 2.4 explains how to calculate the far-field intensity of the secondary radiation. In Section 3, we illustrate this theoretical approach by several examples dealing with fluorescence recorded in different experimental conditions and with different models for the dipole distribution.

**2. Oscillating dipole theory**

## 2.1 Radiation of an oscillating dipole within a periodic multilayer structure

The geometry of the problem is given in Fig. 1. The strategy of the calculation is to find the lateral components of the electric and magnetic fields generated by an oscillating electric dipole that are continuous at the interfaces according to the standard boundary conditions of optics and to propagate these components through the structure. Our approach is based on the direct matrix analysis of the radiation emitted by an oscillating dipole embedded in a periodic stratified structure [11]. The total electromagnetic field **E**, **H** associated with a radiating dipole consists in a homogeneous field $\mathbf{E}_0$, $\mathbf{H}_0$ obtained as the solution of the homogenous differential propagation equation deduced from Maxwell's equations plus an inhomogeneous field $\mathbf{E}_i$, $\mathbf{H}_i$ obtained as a particular solution of the inhomogeneous differential equation with a source (dipole) term, that is:

$$\mathbf{E} = \mathbf{E}_0 + \mathbf{E}_i \qquad (2)$$

and

$$\mathbf{H} = \mathbf{H}_0 + \mathbf{H}_i \qquad (3)$$

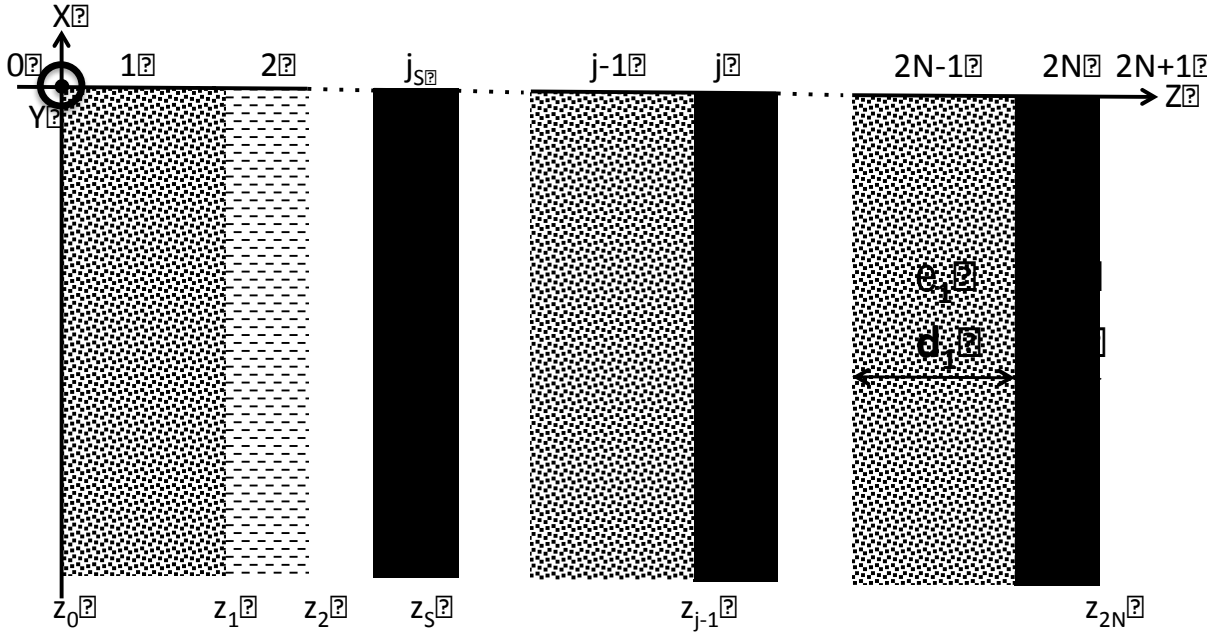

Fig. 1: Geometry of a periodic multilayer made up of two alternating layers of dielectric constant and thickness $\varepsilon_1$, $d_1$ and $\varepsilon_2$, $d_2$ respectively. The number of layers is 2N. The dipole **p** is located at the depth $z_s$ in the $j_s^{th}$ layer.

We consider an electric dipole, the moment of which is denoted by **p**, embedded at the location $\mathbf{r_s} = (\boldsymbol{\rho}_s, z_s)$ ($\boldsymbol{\rho}_s$ lateral components and $z_s$ depth component) within the layer $j_s$ of a multilayer structure of dielectric constant ε($\square$) and oscillating at a frequency $\omega_0$. The charge density associated to this dipole is given in the **k**-ω reciprocal space[1] by:

$$\sigma[\mathbf{k}, \omega] = 2\pi i\, \delta[\omega - \omega_0]\, \mathbf{p}.\mathbf{k}\, e^{-i 2\pi \mathbf{k}.\mathbf{r}_s}$$

(4)

while the current density deduced from the charge density by the continuity equation is:

$$\mathbf{J}[\mathbf{k}, \omega] = 2\pi i\, \omega\, \delta[\omega - \omega_0]\, e^{-i 2\pi \mathbf{k}.\mathbf{r}_s}\, \mathbf{p}$$

(5)

Maxwell's equations in the **k**-ω reciprocal space and the above equations for the charge and current densities lead to the following equations[2] for the electric inhomogeneous field $\mathbf{E}_i[\mathbf{k},\omega]$ and the magnetic inhomogeneous field $\mathbf{H}_i[\mathbf{k},\omega]$:

$$\mathbf{E}_i[\mathbf{k}, \omega] = -4\pi\, \delta[\omega - \omega_0]\, e^{-i 2\pi \mathbf{k}.\mathbf{r}_s}\, \frac{\left(\frac{\omega}{c}\right)^2 \mathbf{p} - \frac{(\mathbf{p}.\mathbf{k})\mathbf{k}}{\varepsilon}}{k^2 - \left(\frac{\omega}{c}\right)^2 \varepsilon}$$

(6)

$$\mathbf{H}_i[\mathbf{k}, \omega] = -4\pi\, \delta[\omega - \omega_0]\, e^{-i 2\pi \mathbf{k}.\mathbf{r}_s}\, \frac{\frac{\omega}{c}}{k^2 - \left(\frac{\omega}{c}\right)^2 \varepsilon}\, \mathbf{k} \wedge \mathbf{p}$$

(7)

To apply the continuity boundary conditions, we introduce the partial Fourier transform defined by the following expression:

$$F[\mathbf{k}_p, \omega, z] = \iiint F[\mathbf{r}, t]\, e^{i\omega t}\, e^{-i \mathbf{k}_p.\mathbf{\rho}}\, dt\, d\rho$$

(8)

Taking into account Eqs. [6-8], it comes:

$$\mathbf{E}[\mathbf{k}_p, \omega, z] = -4\pi\, \delta[\omega - \omega_0]\, e^{-i 2\pi \mathbf{k}_p.\mathbf{r}_s}\, \mathcal{E}[\mathbf{k}, \omega, z]$$

(9)

and

$$\mathbf{H}[\mathbf{k}_p, \omega, z] = -4\pi\, \omega\, \delta[\omega - \omega_0]\, e^{-i 2\pi \mathbf{k}_p.\mathbf{r}_s}\, \mathcal{H}[\mathbf{k}, \omega, z]$$

(10)

where

---

[1] The **k**-$\omega$ domain is the reciprocal space of the **r**-*t* (direct 3D space-time) domain; both domains are mathematically related by the Fourier and inverse Fourier transforms.
[2] All calculations are performed in the Gauss unit system.

$$E[\mathbf{k},\omega,z] = \int \frac{\left(\frac{\omega}{c}\right)^2 \mathbf{p} - \frac{1}{\varepsilon}[(\mathbf{p}.\mathbf{k})\mathbf{k}]}{k^2 - \left(\frac{\omega}{c}\right)^2 \varepsilon} e^{-i 2\pi k_z (z-z_s)} dk_z$$

(11)

and

$$H[\mathbf{k},\omega,z] = \int \frac{\mathbf{k} \wedge \mathbf{p}}{k^2 - \left(\frac{\omega}{c}\right)^2 \varepsilon} e^{-i 2\pi k_z (z-z_s)} dk_z$$

(12)

The above integrals over $k_z$ can be calculated by means of Cauchy theorem.

To treat the propagation of the fields it is convenient to introduce a canonical reference system (X, Y, Z). In this system, which depends on the tangential component of the wave-vector, the tangential component of the field has only one component (say Y, the X component being null). The canonical system can be built as follows: the unit vector Z is along the direction normal to the stratification planes, the unit vector of the Y axis is collinear with the tangential component $\mathbf{k}_\rho$ □□□□□□□□□□□□□□ and the unit vector along X is obtained from the cross product.

Indeed the continuous quantities at the boundaries are the tangential components of the total electric **E** and magnetic field **H**; introducing the quadri-vector $Q[\mathbf{k}, \omega, z, z_s]$ built from the tangential components of **E** and **H** in the canonical system as:

$$Q[\mathbf{k},\omega,z,z_s] = \begin{pmatrix} H_X[\mathbf{k},\omega,z,z_s] \\ E_Y[\mathbf{k},\omega,z,z_s] \\ E_X[\mathbf{k},\omega,z,z_s] \\ H_Y[\mathbf{k},\omega,z,z_s] \end{pmatrix}$$

(13)

Then boundary conditions applied at the interface between the layer *j-1* and the layer *j*, located at $z_{j-1}$ lead to the set of recurrent equation:

$$Q_{j-1}[\mathbf{k},\omega,z_{j-1},z_s] = Q_j[\mathbf{k},\omega,z_{j-1},z_s]$$

(14)

At this stage two cases have to be considered:

- the layer *j* does not contain the dipole; in this case one has:

$$Q_j[\mathbf{k},\omega,z_j,z_s] = Q_{0j}[\mathbf{k},\omega,z_j]$$

(15)

where $Q_{0j}$ is the quadri-vector equivalent to $Q_j$ but constructed from the tangential components of the homogeneous fields, that is:

$$Q_0[\mathbf{k},\omega,z] = \begin{pmatrix} H_{0X}[\mathbf{k},\omega,z] \\ E_{0Y}[\mathbf{k},\omega,z] \\ E_{0X}[\mathbf{k},\omega,z] \\ H_{0Y}[\mathbf{k},\omega,z] \end{pmatrix} \quad (16)$$

- the layer $j$ does contain the dipole; in this case one has:

$$Q_j[\mathbf{k},\omega,z_j,z_s] = Q_{0j}[\mathbf{k},\omega,z_j] + S[\mathbf{k},\omega,z_j,z_s] e^{-i k_z |z_j - z_s|} \quad (17)$$

since the total field must include the source term S; by performing the integration of Eqs. (11) and (12), one finds:

$$S[\mathbf{k},\omega,z,z_s] = \begin{pmatrix} -2\pi i\, k_o\, p_Y\, sgn[z-z_s] \\ -2\pi i\, k_o^2\, \dfrac{p_Y}{k_z} \\ -\dfrac{2\pi i}{\varepsilon}(k_z p_x + k_p p_z\, sgn[z-z_s]) \\ 2\pi i\, k_o\left(\dfrac{k_p p_z}{k_z} + p_x\, sgn[z-z_s]\right) \end{pmatrix} \quad (18)$$

The homogeneous quadri-vector $Q_{0j-1}$ at the interface $j-1$ can be deduced from the quadri-vector at the interface $j-2$ by means of the formalism given by Abelès [15,16], that is:

$$Q_{0j-1}[\mathbf{k},\omega,z_{j-1}] = A[\omega, z_{j-1} - z_{j-2}, \varepsilon_{j-1}] Q_{0j-1}[\mathbf{k},\omega,z_{j-2}] \quad (19)$$

where $A$ is the 4 x 4 transfer Abelès matrix; in the canonical system, this matrix takes the diagonal form in the layer $j$ of thickness $d_j$ and of dielectric constant $\varepsilon_j$:

$$A[\omega, d_j, \varepsilon_j] = \begin{pmatrix} A^{TM}[\omega, d_j, \varepsilon_j] & 0 \\ 0 & A^{TE}[\omega, d_j, \varepsilon_j] \end{pmatrix} \quad (20)$$

$A^{TM}$ and $A^{TE}$ are the 2 x 2 Abelès matrices for the Transverse Magnetic (TM) and Transverse Electric (TE) polarizations respectively, whose expressions are given in [17]. Note that the matrix A does not depend on the layer number but only on its nature (layer of kind 1 or 2). Each field can be split into a transmitted $T$ and reflected $R$ component so that finally the electromagnetic field is given in the so-called $T$-$R$ representation by a quadri-vector $TR$:

$$TR[\mathbf{k},\omega,z,z_s] = \begin{pmatrix} T_{TM}[\mathbf{k},\omega,z,z_s] \\ R_{TM}[\mathbf{k},\omega,z,z_s] \\ T_{TE}[\mathbf{k},\omega,z,z_s] \\ R_{TE}[\mathbf{k},\omega,z,z_s] \end{pmatrix} \quad (21)$$

In the canonical system, the quadri-vector *TR* transforms into the homogeneous quadri-vector $Q_0$ by means of the 4 x 4 matrix *M*:

$$Q_0 = M\, TR$$

(22)

with

$$M[k_\parallel, k_z] = \begin{pmatrix} \frac{-k_z}{k_0} & \frac{1}{k_z} & 0 & 0 \\ \frac{1}{k_0} & \frac{1}{k_\parallel} & 0 & 0 \\ 0 & 0 & 1 & 1 \\ 0 & 0 & \frac{-k_z}{k_0} & \frac{k_z}{k_\parallel} \end{pmatrix}$$

(23)

Now we have implemented all the tools required to calculate the field amplitudes in the far-field region. Hereafter we present the algorithm to perform it. Let us assume that the dipole is located in the layer *2m-1*; see Figure 1 for the geometry. The calculation is driven in five main steps:

1. one starts from the amplitude $R_{TM}$ and $R_{TE}$ of the fields at the interface $z_0$ between the external medium *j=0* and the first layer of the stack *j=1*; since one assumes that there is no incoming wave from the external medium *j=0*, then $T^{TM}=0$ and $T^{TE}=0$ and

$$Q_{00}[\mathbf{k}, \omega, z_0] = M[k_0, k_{z0}] \begin{pmatrix} 0 \\ R_{TM}[\mathbf{k}, \omega, z_0, z_s] \\ 0 \\ R_{TE}[\mathbf{k}, \omega, z_0, z_s] \end{pmatrix}$$

(24)

2. one propagates the field up to the interface $z_{2m-2}$ by means of the 4 x 4 Abelès matrices, that is:

$$Q_{02m-2}[\mathbf{k}, \omega, z_{2m-2}] = (A[\omega, d_2, \varepsilon_2] A[\omega, d_1, \varepsilon_1])^{m-1} Q_{00}[\mathbf{k}, \omega, z_0]$$

(25)

3. one propagates the field through the bilayer containing the dipole to get the field $Q_{02m-2}[\mathbf{k}, \omega, z_{2m}]$; the procedure is summarized in the Appendix.

4. one continues the propagation up to the external layer *j=2 N+1*:

$$Q_{02N+1}[\mathbf{k}, \omega, z_{2N+1}] = (A[\omega, d_2, \varepsilon_2] A[\omega, d_1, \varepsilon_1])^{n-m} Q_{02m-1}[\mathbf{k}, \omega, z_{2m}]$$

(26)

5. finally from the field $Q_{02N+1}[\mathbf{k}, \omega, z_{2N+1}]$ one deduces the amplitude $T_{TM}[\mathbf{k}, \omega, z_{2N+1}, z_s]$ and $T_{TE}[\mathbf{k}, \omega, z_{2N+1}, z_s]$ at the interface $z_{2N+1}$, taking into account that there is no incoming wave in the external medium *j=2N+1*:

$$\begin{pmatrix} T_{TM}[\mathbf{k}, \omega, z_{2N+1}, z_s] \\ 0 \\ T_{TE}[\mathbf{k}, \omega, z_{2N+1}, z_s] \\ 0 \end{pmatrix} = M^{-1}[k_0, k_{z2N}] Q_{u2N+1}[\mathbf{k}, \omega, z_{2N+1}]$$

(27)

Combining the equations (24-27) leads to a system of four equations whose unknowns are: $T_{TM}[\mathbf{k}, \omega, z_{2N+1}, z_s]$, $T_{TE}[\mathbf{k}, \omega, z_{2N+1}, z_s]$, $R_{TM}[\mathbf{k}, \omega, z_0, z_s]$ and $R_{TE}[\mathbf{k}, \omega, z_0, z_s]$.

Solving this system provides the amplitudes of the field radiated in the external media. In practice the radiation is detected in the far-field region at a distance larger than the wavelength. The way to calculate the intensity of secondary X-rays in far-field is detailed in Section 2.3.

## 2.2 Calculation of the in-depth distribution of the local electric field

As mentioned previously, the amplitude of the electric field that excites the dipole is generally computed via "rigorous" methods such as Parratt recursive method or transfer matrix technique. Nevertheless the coupled-wave theory (CWT) appears in terms of computing time to be very efficient to deal with this problem since the field can be expressed by a simple formula as shown hereafter; it is the reason why one chooses to use this approach. In the TE polarization case, the electric field has only one component $E(z)$ which is perpendicular to the incident plane and obeys the wave equation:

$$\frac{d^2 E(z)}{dz^2} + \varepsilon(z) k^2 E(z) = 0$$

(28)

In the CWT, E($z$) is regarded as a superposition of two waves propagating in opposite directions along the $z$-axis with amplitude F (Forward) and B (Backward) varying with $z$ ; one writes:

$$E(z) = F(z)e^{ik_z z} + B(z)e^{-ik_z z}$$

(29)

To ensure a unique determination of the amplitude, one requires the following condition to be satisfied:

$$F'(z)e^{ik_z z} + B'(z)e^{-ik_z z} = 0$$

(30)

the prime symbol indicating a derivative with respect to $z$. Combining these two equations leads to the following system of differential equations:

$$F'(z) = -\frac{i\,k^2}{2\,k_{z0}}\,\chi(z)\,\left(F(z) + B(z)e^{-i\,2\,k_z\,z}\right)$$

$$B'(z) = +\frac{i\,k^2}{2\,k_{z0}}\,\chi(z)\,\left(F(z)e^{+i\,2\,k_z\,z} + B(z)\right)$$

(31)

together with the boundary conditions F(z=0)=1 and B(z=z$_{2N+1}$=L)=0.

The quantity $\chi(z) = 1 - \varepsilon(z)$ can be written by means of the piecewise function h(z):

$$\chi(z) = \bar{\chi} + (\chi_2 - \chi_1)\,h(z)$$

(32)

where $\bar{\chi}$ is the average value of $\chi(z)$; h(z) can be expanded in Fourier series:

$$h(z) = \sum_{n=-\infty}^{+\infty} h_n \exp\left[\frac{2\,i\,\pi\,n}{d}\,z\right]\,;\,h_n = \frac{1}{2\,i\,\pi\,n}\,(1 - \exp[-2\,i\,\pi\,n\,\gamma])$$

(33)

Combining the previous equations and assuming that the multilayer diffracts the incident radiation at the $p^{th}$ order, that is the Bragg condition $k_z = \frac{n\,\pi}{d}$ is nearly satisfied, then a system of coupled differential equations with constant coefficients can be obtained for the quantities:

$$f(z) = F(z)\,\exp\left[-i\,\left(\frac{n\,\pi}{d} - k_{z0}\right)\,z\right]$$

(34)

and

$$b(z) = B(z)\,\exp\left[+i\,\left(\frac{n\,\pi}{d} - k_{z0}\right)\,z\right]$$

(35)

The system reads:

$$f'(z) + \alpha\,f(z) + \beta\,b(z) = 0$$
$$b'(z) - \alpha\,b(z) + \gamma\,f(z) = 0$$

(36)

where

$$\alpha = i\left(\frac{n\,\pi}{d} - k_{z0} + \bar{\chi}\frac{k^2}{2\,k_{z0}}\right)$$

(37)

$$\beta = i\,\frac{k^2}{2\,k_{z0}}\,(\chi_2 - \chi_1)\,h_n$$

(38)

$$\gamma = -i\frac{k^2}{2k_{z0}}(\chi_2 - \chi_1)h_{-n}$$

(39)

with the boundary conditions f(z=0)=1 and b(z=z$_{2N+1}$=L)=0. Solving the system of Eqs. (35-39) gives:

$$f(z) = \frac{r\,\mathrm{Cosh}[r(L-z)] + \alpha\,\mathrm{Sinh}[r(L-z)]}{r\,\mathrm{Cosh}[rL] + \alpha\,\mathrm{Sinh}[rL]}$$

(40)

and

$$b(z) = \frac{\gamma\,\mathrm{Sinh}[r(L-z)]}{r\,\mathrm{Cosh}[rL] + \alpha\,\mathrm{Sinh}[rL]}$$

(41)

with

$$r = \sqrt{\alpha^2 + \beta\gamma}$$

(42)

The above calculation given for the TE polarization case can be transposed to the TM case by considering the wave equation governing the magnetic field instead of the electric one. Indeed in the TM polarization case, the magnetic field has only one component $H(z)$ which is perpendicular the incident plane:

$$\frac{d^2H(z)}{dz^2} - \frac{d\ln[\varepsilon(z)]}{dz}\frac{dH(z)}{dz} + \varepsilon(z)k^2H(z) = 0$$

(43)

To simplify this equation one can introduce the field $H^*(z)$:

$$H^*(z) = \frac{H(z)}{\varepsilon^*(z)}$$

(44)

where

$$\varepsilon^*(z) = \varepsilon(z) + \frac{1}{2k^2}\frac{1}{\varepsilon(z)}\frac{d^2\varepsilon(z)}{dz^2} - \frac{3}{4k^2}\left(\frac{1}{\varepsilon(z)}\frac{d^2\varepsilon(z)}{dz^2}\right)^2$$

(45)

so that $H^*(z)$ satisfies the wave equation:

$$\frac{d^2H^*(z)}{dz^2} + \varepsilon^*(z)k^2H^*(z) = 0$$

(46)

The calculation to get $H^*(z)$ is formally the same than the one given above to determine the electric field $E(z)$ in the TE polarization case.

## 2.3 Calculation of the induced dipole

We assume that the media are linear and isotropic, so that the polarization is aligned with and proportional to the local electric field **E** at position of the dipole $\mathbf{r}_s$:

$$p(\omega_0,\mathbf{r}_s) = \chi(\omega_0,\mathbf{r}_s)E(\omega_0,\mathbf{r}_s)$$

(47)

where $\chi(\omega_0,\mathbf{r}_s)$ is the electric susceptibility of the medium. If the medium is not very dispersive the dependence of the susceptibility on the frequency $\omega_0$ can be discarded; otherwise, for instance in the vicinity of an absorption edge, it can be necessary to take into account the dispersion. This is not a simple talk. In first attempt, one can call upon the Clausius-Mossotti formula [18].

To perform simulation, it is necessary to model the in-depth distribution of the dipoles. In the following part of this paper, we consider two models:

- the first one where the scatterers are uniformly distributed along each layer; a fraction *f* is in the layer, say 1, while the remaining *1-f* is in the other layer, say 2.
- the second one describes the situation where the interfaces are not sharp and a transition layer is formed at each interface. To model this case, we call upon the error function (erf) as it is usually done to model rough interfaces [4,19]. Let us emphasize that the transition layer is not necessary the same for a material *a* on the top of a material *b* than for a material *b* on the top of a material *a;* see for instance [20–22]. In this case, we use the following distribution profile function *p(z)*:

$$p(z) = \sum_{h}^{Nb} p1(z,h)\,\theta[z - h\,d]\,\theta[\gamma\,d - (z - h\,d)] + p2(z,h)\,\theta[d - (z - h\,d)]\,\theta[-\gamma\,d + (z - h\,d)]$$

(48)

☐ ☐ ☐ ☐ θ being the unit step Heaviside function and

$$p1(z,h) = 0.5\,Erf\left[\frac{z - hd}{\sqrt{2}\,\sigma 1}\right] + 0.5\,Erf\left[\frac{\gamma\,d - (z - hd)}{\sqrt{2}\,\sigma 2}\right]$$

(49)

and

$$p2(z,h) = 1 - \left( 0.5\, Erf\left[\frac{(z-hd) - \gamma d}{\sqrt{2}\,\sigma 1}\right] + 0.5\, Erf\left[\frac{d - (z-hd)}{\sqrt{2}\,\sigma 2}\right] \right)$$

(50)

This profile gives a distribution of the dipoles close to 1 in the centre of the layer 1, close to 0 in the middle of the layer 2 with diffuse interfaces (1–2 between layer 1 and 2, and 2–1 between layer 2 and 1). The parameters σ1 and σ2 can be regarded as the *rms* roughness of the interfaces 1-2 and 2-1 respectively. Discussion concerning the relationship between interface roughness, diffuse interface and transition layer can be found in [19,23]. Nevertheless as mentioned by Ghose and Dev [4], the values σ1 and σ2 may be inconsistent with the *rms* roughness values obtained from X-ray reflectometry fit.

**2.4 Radiated intensity in the far-field**

The problem is to calculate the intensity of radiation detected in a direction given by **D**=($\rho_D$, $z_D$) at a distance large from the multilayer structure. The first step consists in determining the field F[**r**=**D**,ω] in the **r** (direct 3D space) domain (**F** being the electric or magnetic field) from the field **F**[$\mathbf{k}_\rho$,ω, z] as calculated in section 2.1; this operation requires an integration over the parallel component $\mathbf{k}_\rho$ of a kernel containing the field **F**[$\mathbf{k}_\rho$,ω, z]

$$F[\mathbf{D},\omega] = \iint d\mathbf{k}_\rho\, F[\mathbf{k}_\rho, \omega, z_0]\, \exp[i\,\mathbf{k}_\rho\,\boldsymbol{\rho}_D]\, \exp[i\,\mathbf{k}_\perp(\mathbf{k}_\rho,\omega)\,z_D]\, \frac{\partial k_\perp(\mathbf{k}_\rho,\omega)}{\partial \omega}$$ (51)

The calculation can be done by using the stationary phase method (SPM) as done in [13,24] : roughly speaking, the integration by the SPM results in replacing $\mathbf{k}_\rho \to \mathbf{k}_D$ the lateral reciprocal lateral wave-vector associated with $\boldsymbol{\rho}_D$ and to multiply by the result by $cos(\theta)$ and by constant terms which are irrelevant when no absolute value is looking for. A detailed mathematical development is given in Ref. [25]. The second step corresponds to take the squared modulus of this result.

**3. Numerical applications**

In a first step, we compare our theory with data published in the literature in the case where only the incident radiation undergoes Bragg diffraction (de Boer Mode: dBM). In a second step we consider the case where both primary and secondary radiations are diffracted in the Bragg condition (Kossel-Chauvineau-Bridou Mode: KCBM).

First we consider the case of the Pt/C multilayer reported by Ghose and Dev [4]. The structure consists in 20 bilayers; the thickness of the Pt layer is 1.7 nm and the one of the C layer is 2.6 nm. The L fluorescence line of Pt (9400 eV) is excited by the Mo K$\alpha$1 radiation (17487.36 eV); the average exit angle is 50°. Figure 2 shows the Pt L fluorescence yield recorded in the dBM, versus the glancing angle computed by means of our model in absence of any roughness. It appears that our calculation is in agreement with the data reported in [4].

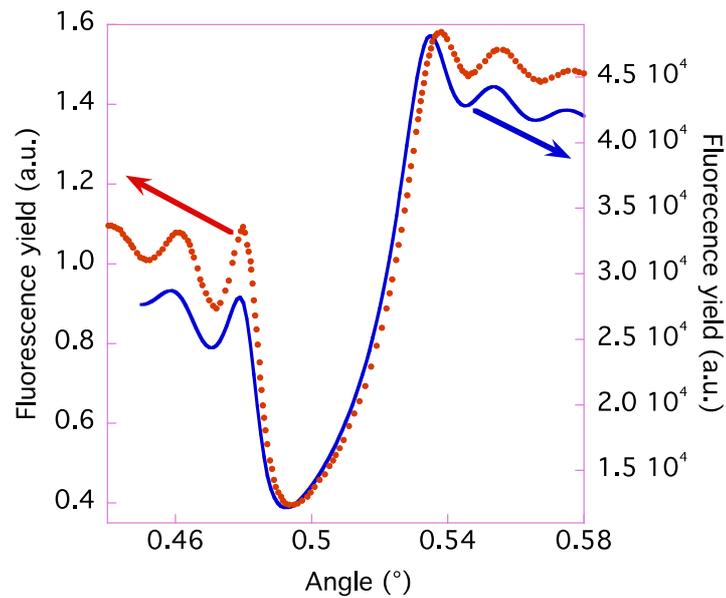

Fig. 2: Pt L fluorescence yield of a Pt/C multilayer from Ref. [4] (red dots) compared to our calculation (blue line).

As a second example in the dBM, we consider the Mo/Si multilayer system studied by Tiwari and Sawhney [5]. The structure has 20 bilayers; the thickness of the Mo layer is 2.376 nm and the one of the Si layer is 4.224 nm. The L$\alpha$ fluorescence line of Mo (2293 eV) is excited by the 15 keV monochromatic radiation delivered on the B16 beamline at the Diamond Light Source and recorded with an average exit angle of 90°. Our result presented in Figure 3 can be compared to the values given in figures 5 and 8 of the referenced paper.

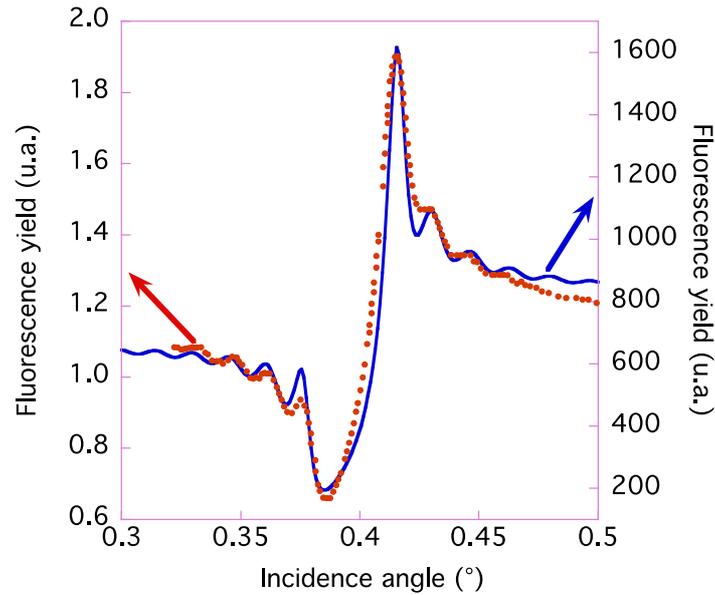

Fig. 3: Mo L fluorescence yield of a Mo/Si multilayer from Ref. [5] (red dots) compared to our calculation (blue line). Our calculation has been shifted by +0.0025°.

Let us now consider the KCBM as reported in [2,3]. The sample is a Fe/C multilayer with 24 bilayers; the thickness of the Fe layer is 2.80 nm and the one of the C layer is 2.56 nm. The K fluorescence line of Fe (6404 eV) is excited with the Kα line of Cu (8084 eV) Bragg diffracted by the periodic arrangement of the multilayer (glancing angle $\theta_0$=0.88°). Figure 4 shows the fluorescence yield versus the take-off angle in the vicinity of the Bragg angle for the Fe K line (about 1.1°) as calculated by means of our model in comparison with the experimental data [2,3]. In our model, the diffraction within the multilayer stack is calculated by taking into account the interface roughness with the *rms* values given in [2,3] but no inter-diffusion is considered. The discrepancy between the two curves can be attributed to several factors: the interfaces are likely diffuse; broadening factors are not included in the calculation; complicated problems of radiation polarization are not taken into account and the geometry with problem of solid angle of detection and footprint of the beam.

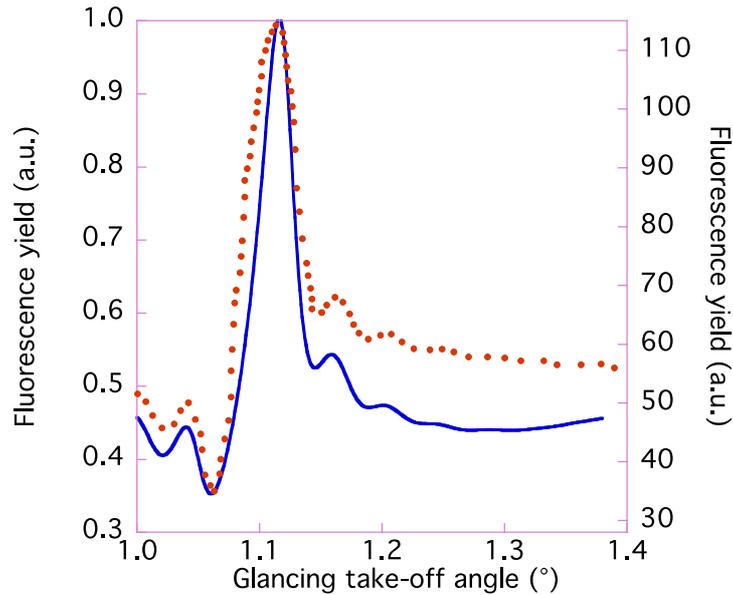

Fig. 4: Fe K fluorescence yield of a Fe/C multilayer from Ref. [2,3] (red dots) compared to our calculation (blue line).

With our model, it is possible to simulate the effect of inter-diffusion. Let us deal with two cases: inter-diffusion with a uniform distribution along the layer and diffuse interfaces. We consider the Fe/C multilayer sample of the previous example. Figure 5 illustrates the effect of uniform mixing: the fraction $f$ is 0.1 and 0.2 which means that 10 and 20% of Fe atoms are uniformly distributed in the C layer, respectively; the ideal case (no inter-diffusion) is shown for comparison. The yields have been normalized with respect to their maximum. As a function of increasing $f$, the first dip toward the low angles shifts by +0.005° and its intensity increases from 33 to 45% of the intensity of the main peak. Accurate measurements using synchrotron radiation should make possible to record the effects of mixing larger than 10 % but this task seems more difficult to achieve with a laboratory experiment equipped with an X-ray tube.

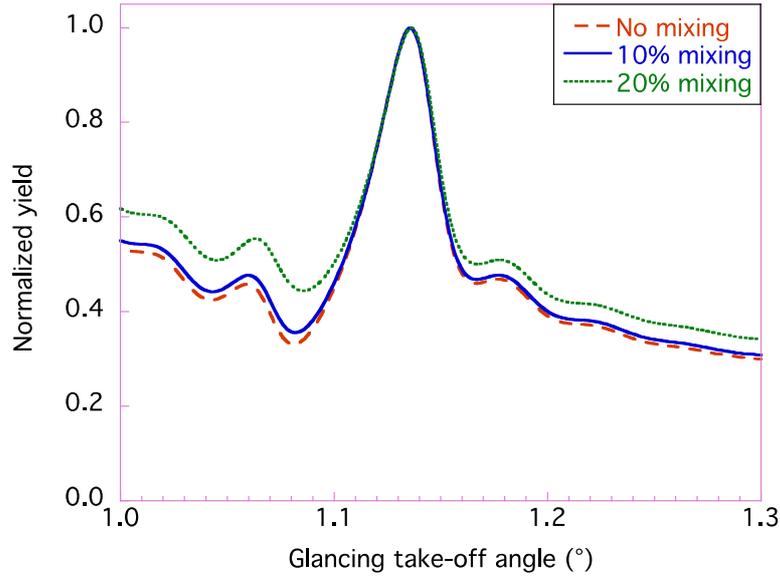

Figure 5: Effect of uniform mixing on the Fe K fluorescence yield of a Fe/C multilayer defined in Ref. [2,3]. Blue solid line: the fraction $f$ is 0.1; green dotted line: $f$ is 0.2; red dashed line: ideal case.

Figure 6 shows the effect of diffuse interfaces; the profile is modelled by Eqs. (48-50). The parameters are $\sigma 1=0.8$ nm and $\sigma 2=0.2$ nm. We note that the changes are very small and may be difficult to observe even with synchrotron radiation.

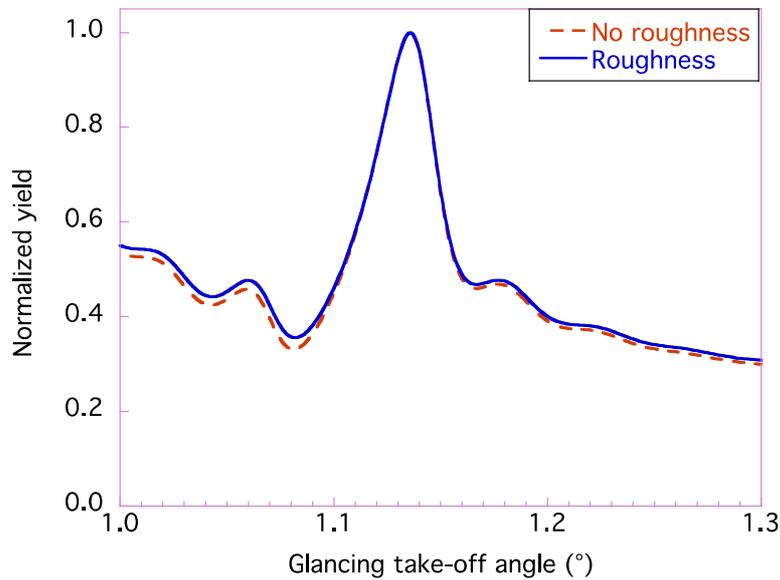

Fig. 6: Effect of diffuse interfaces on the Fe K fluorescence yield of a Fe/C multilayer defined in Ref. [2,3]. Blue solid line: diffuse interface with $\sigma 1=0.8$ nm and $\sigma 2=0.2$ nm; red dashed line: ideal case.

It is not the purpose of the present paper to discuss the sensitivity of the fluorescence yield to the distribution of the scattering elements; this has been done in [4] where it was suggested that X-ray reflectometry should be combined with X-ray standing waves to access microstructural details of periodic multilayers. Let us mention that interesting results using simultaneous analysis of X-ray grazing incidence reflectivity and angular dependent fluorescence from ultrathin La films were recently obtained in layered La/B$_4$C structure forming a waveguide [6]. Although this waveguide does not work as a Bragg reflector, it is important to outline that our dipole model can be implemented to analyse this waveguide structure. Finally let us outline that the model is in principle valid in the domain of total reflection; nevertheless we have observed some numerical instabilities in this region that we were not able to overcome in our code. This problem is under study.

## 4. Conclusion

We have developed a model for the standing wave enhanced X-ray fluorescence that allows one to take into account the Bragg diffraction of both primary and secondary radiations. This model can be also useful to analyse other layered systems such as waveguide supporting several waveguide modes. Its in the soft X-ray domain, where absorption cannot be neglected, is another interest of this approach. We have also shown that the coupled-wave theory appears to be an efficient method in terms of computing time, to calculate the in-depth distribution of the exciting electric field. Finally let us mention that the model can be extended to deal with grazing exit fluorescence experiments [26–28] or elastic (Rayleigh) and inelastic (Compton, Raman). Physically, this model is valid provided that the emission process can be treated in the framework of the classical model of the oscillating dipole.

***Acknowledgments****: Dr. F. Bridou from Institut d'Optique in Palaiseau, France, is thanked for helpful discussions.*

**References**

[1] D.K.G. de Boer, Glancing-incidence x-ray fluorescence of layered materials, Phys. Rev. B. 44 (1991) 498-511.
[2] J.-P. Chauvineau, F. Bridou, Analyse angulaire de la fluorescence du fer dans une multicouche périodique Fe/C, J. Phys. IV. 06 (1996) C7-53-C7-64 (in French).
[3] F. Bridou, J.-P. Chauvineau, A. Mirone, Étude de la fluorescence du fer dans une multicouche périodique Fe/C éclairée sous incidence rasante par un faisceau de rayons X monochromatique, J. Phys. IV. 08 (1998) Pr5-309-Pr5-316 (in French).
[4] S.K. Ghose, B.N. Dev, X-ray standing wave and reflectometric characterization of multilayer structures, Phys. Rev. B. 63 (2001) 245409.
[5] M.K. Tiwari, K.J.S. Sawhney, Structural characterization of thin layered materials using x-ray standing wave enhanced elastic and inelastic scattering measurements, J. Phys. Condens. Mat. 22 (2010) 175003.
[6] I.A. Makhotkin, E. Louis, R.W.E. van de Kruijs, A.E. Yakshin, E. Zoethout, A.Y. Seregin, et al., Determination of the density of ultrathin La films in La/B4C layered structures using X-ray standing waves, phys. stat. sol. (b). 208 (2011) 2597-2600.
[7] L.G. Parratt, Surface Studies of Solids by Total Reflection of X-Rays, Phys. Rev. 95 (1954) 359-369.
[8] D.L. Windt, IMD—Software for modeling the optical properties of multilayer films, Comp. Phys. 12 (1998) 360-370.
[9] W. Kossel, V. Loeck, H. Voges, Die Richtungsverteilung der in einem Kristall entstandenen charakteristischen Röntgenstrahlung, Z. Physik. 94 (1935) 139-144 (in German).
[10] P. Jonnard, J.-M. André, C. Bonnelle, F. Bridou, B. Pardo, Soft-x-ray Kossel structures from W/C multilayers under various electron ionization conditions, Phys. Rev. A. 68 (2003) 032505.
[11] J.-M. André, P. Jonnard, B. Pardo, Radiation emitted by an oscillating dipole embedded in a periodic stratified structure: A direct matrix analysis, Phys. Rev. A. 70 (2004) 012503.
[12] M. von Laue, Die Fluoreszenzröntgenstrahlung von Einkristallen (Mit einem Anhang über Elektronenbeugung), Annal. Phys. 415 (1935) 705–746 (in German).
[13] C.E. Reed, J. Giergiel, J.C. Hemminger, S. Ushioda, Dipole radiation in a multilayer geometry, Phys. Rev. B. 36 (1987) 4990-5000.
[14] S. Barkeshli, P.H. Pathak, On the dyadic Green's function for a planar multilayered dielectric/magnetic media, IEEE Trans. Microwave Theo. Techn. 40 (1992) 128-142.
[15] F. Abelès, Recherches sur la propagation des ondes électromagnétiques sinusoidales dans les milieux stratifiés: application aux couches minces, Annal. Phys. (Fr.). 5 (1950) 596-640 (in French).
[16] F. Abelès, La théorie générale des couches minces, Journal de Physique et le Radium. 11 (1950) 307-309 (in French).
[17] M. Born, E. Wolf, Principles of Optics: Electromagnetic Theory of Propagation, Interference and Diffraction of Light, 4th éd., Cambridge University Press, 1970, Section 1.6.2.
[18] J.D. Jackson, Classical Electrodynamics, 2e éd., Wiley, 1975, Section 4.5, p 154-158.
[19] J. Daillant, A. Gibaud, éd., X-ray and Neutron Reflectivity: Principles and Applications, 1re éd., Springer, 2008, Chapters 2, 4 and 6.
[20] S. Yulin, T. Feigl, T. Kuhlmann, N. Kaiser, A.I. Fedorenko, V.V. Kondratenko, et al.,

Interlayer transition zones in Mo/Si superlattices, J. Appl. Phys. 92 (2002) 1216-1220.

[21]  H. Maury, P. Jonnard, J.-M. André, J. Gautier, F. Bridou, F. Delmotte, et al., Interface characteristics of Mo/Si and B4C/Mo/Si multilayers using non-destructive X-ray techniques, Surf. Sci. 601 (2007) 2315-2322.

[22]  H. Maury, J.-M. André, K. Le Guen, N. Mahne, A. Giglia, S. Nannarone, et al., Analysis of periodic Mo/Si multilayers: Influence of the Mo thickness, Surf. Sci. 603 (2009) 407-411.

[23]  B. Pardo, T. Megademini, J.-M. André, X-UV synthetic interference mirrors : theoretical approach, Rev. Phys. Appl. 23 (1988) 1579-1597.

[24]  B. Laks, D.L. Mills, Photon emission from slightly roughened tunnel junctions, Phys. Rev. B. 20 (1979) 4962-4980.

[25]  B. Pardo, J.-M. André, Classical theory of resonant transition radiation in multilayer structures, Phys. Rev. E. 63 (2000) 016613.

[26]  K. Tsuji, Grazing-exit electron probe X-ray microanalysis (GE-EPMA): Fundamental and applications, Spectrochimica Acta Part B: Atomic Spectroscopy. 60 (2005) 1381-1391.

[27]  J. Yang, K. Tsuji, D. Han, X. Ding, GE-MXRF analysis of multilayer films, X-Ray Spectrom. 37 (2008) 625–628.

[28]  T. Awane, S. Fukuoka, K. Nakamachi, K. Tsuji, Grazing exit micro x-ray fluorescence analysis of a hazardous metal attached to a plant leaf surface using an x-ray absorber method, Anal. Chem. 81 (2009) 3356-3364.

**Appendix**

We give the algorithm to calculate the homogeneous field $Q_0[\mathbf{k},\omega,z]$ in the $j=2m+1$ layer from the homogeneous field in the layer $j=2m-2$, assuming that the dipole is located at $z_s$ within the layer $j=2m-1$. By continuity and absence of dipole in the layer $j=2m-2$, Eqs. (14) and (15) give:

$$Q_{2m-1}[\mathbf{k},\omega,z_{2m-2},z_s] = Q_{02m-2}[\mathbf{k},\omega,z_{2m-2}]$$

(A.1)

Since the dipole is in the layer $j=2m-1$, from Eq. [17] it comes:

$$Q_{02m-1}[\mathbf{k},\omega,z_{2m-2}] = Q_{2m-1}[\mathbf{k},\omega,z_{2m-2},z_s] - S[\mathbf{k},\omega,z_{2m-1},z_s]\, e^{-i k_z |z_{2m-2}-z_s|}$$

(A.2)

Propagating the field in the layer $j=2m-1$, one gets:

$$Q_{02m-1}[\mathbf{k},\omega,z_{2m-1}] = A[\omega,z_{2m-1}-z_{2m-2},\varepsilon_{2m-1}]Q_{02m-1}[\mathbf{k},\omega,z_{2m-2}]$$

(A.3)

Applying Eq. (17) since the dipole is in the layer $2m-1$, one has:

$$Q_{2m-1}[\mathbf{k},\omega,z_{2m-1},z_s] = Q_{02m-1}[\mathbf{k},\omega,z_{2m-1}] + S[\mathbf{k},\omega,z_{2m-1},z_s]\, e^{-i k_z |z_{2m-1}-z_s|}$$

(A.4)

Eqs. (14) and (15) give by virtue of the continuity of the field and absence of dipole in the layer $j=2m$:

$$Q_{02m}[\mathbf{k},\omega,z_{2m-1}] = Q_{2m}[\mathbf{k},\omega,z_{2m-1},z_s]$$

(A.5)

Propagating the field in the layer $j=2m$ by means of Eq. (19) leads to:

$$Q_{02m+1}[\mathbf{k},\omega,z_{2m}] = A[\omega,z_{2m}-z_{2m-1},\varepsilon_{2m}]Q_{02m}[\mathbf{k},\omega,z_{2m-1}]$$

(A.6)